\documentclass[aps,prb,twocolumn,groupedaddress]{revtex4}

\newif\ifpdf
\ifx\pdfoutput\undefined
\pdffalse % we are not running PDFLaTeX
\else
\pdfoutput=1 % we are running PDFLaTeX
\pdftrue
\fi

\ifpdf
\usepackage[pdftex]{graphicx}
\else
\usepackage{graphicx}
\fi
\usepackage{hyperref}

\begin{document}

\ifpdf
\DeclareGraphicsExtensions{.pdf, .jpg}
\else
\DeclareGraphicsExtensions{.eps, .jpg}
\fi

\def\hslash{\hbar}
\def\imag{i}
\def\grad{\vec{\nabla}}
\def\div{\vec{\nabla}\cdot}
\def\curl{\vec{\nabla}\times}
\def\DDt{\frac{d}{dt}}
\def\ddt{\frac{\partial}{\partial t}}
\def\ddx{\frac{\partial}{\partial x}}
\def\ddy{\frac{\partial}{\partial y}}
\def\lap{\nabla^{2}}
\def\divv{\vec{\nabla}\cdot\vec{v}}
\def\gradS{\vec{\nabla}S}
\def\vvec{\vec{v}}
\def\wc{\omega_{c}}
\def\<{\langle}
\def\>{\rangle}
\def\Tr{{\rm Tr}}
\def\Csch{{\rm csch}}
\def\Coth{{\rm coth}}
\def\Tanh{{\rm tanh}}
\def\g2{g^{(2)}}
\newcommand{\al}{\alpha}

\newcommand{\la}{\lambda}
\newcommand{\del}{\delta}
\newcommand{\om}{\omega}
\newcommand{\ep}{\epsilon}
\newcommand{\pd}{\partial}
\newcommand{\bra}{\langle}
\newcommand{\ket}{\rangle}
\newcommand{\bbra}{\langle \langle}
\newcommand{\kket}{\rangle \rangle}
\newcommand{\non}{\nonumber}
\newcommand{\be}{\begin{equation}}
\newcommand{\ee}{\end{equation}}
\newcommand{\bea}{\begin{eqnarray}}
\newcommand{\eea}{\end{eqnarray}}

\title{Constraints on the two-particle distribution function due to the 
permutational symmetry of the higher
order distribution functions}

\author{Andrey Pereverzev}
\email{aperever@mail.uh.edu}
\affiliation{Department of Chemistry and Center for Materials Chemistry, 
University of Houston \\ Houston, TX 77204}
\date{\today}

\begin{abstract}
We investigate how the range of parameters that specify the two-particle distribution 
function is restricted if we require that this function be obtained 
from the $n^{\rm th}$ order distribution functions that are symmetric
with respect to the permutation of any two particles. We consider the simple case when
 each variable in the distribution functions can take only two values.
Results for all $n$ values are given, including the limit
of $n\to\infty$. We use our results to obtain bounds on the allowed values of 
magnetization and magnetic susceptibility in an $n$ particle Fermi fluid.
\end{abstract}


\maketitle
\section{Introduction}
Two particle coordinate distribution function is one of the central objects of 
statistical mechanics \cite{Balescu, Reichl}.
Any such distribution must satisfy the following requirements. 
Firstly, it must be non-negative
everywhere. Secondly, it must be normalized. Thirdly, for systems of $n$ indistinguishable  
particles, it is obtained by integrating out (or summing out) $n-2$ coordinates of 
the $n$ particle distribution function
that is invariant under the permutation of any of its particles. While the first two requirements
give explicit constrains on the possible forms of two particle distribution functions, this is not the case
for the third requirement. Clearly, it follows from the third requirement that the two particle distribution 
must be symmetric with respect to the permutation of the two particles. However,  the third requirement is
more restrictive, i.e. not all symmetric non-negative and normalized two particle distribution functions 
can be obtained from symmetric $n$ particle distribution functions. Therefore, it is of interest to
understand what are the explicit constraints on the possible forms the two particle distribution function
(other than its symmetry) that follow from the third requirement. This knowledge can be 
useful in various ways.
For example, in the case of equilibrium distributions it would give us an opportunity 
to tell which features of 
the two particle distribution function are due to the permutational symmetry 
of the Hamiltonian and which ones are due to its explicit form. In particular,
as will be shown in a specific example below, symmetry imposed bounds 
on the values of certain physical quantities
can be obtained.   Another field 
of applications includes solving various correlation function integral 
equations \cite{Hansen, Kalikmanov} by  iterational
  procedures . For such equations initial guess function that incorporate 
  the symmetry information can provide  faster convergence.  
 
 Our goal in this paper is two-fold. Firstly, we would like to work out a simple example 
 that shows explicitly how
 the two particle distribution function is restricted by the fact that it is obtained from the 
 permutationally symmetric $n$ particle distribution function. Secondly, we want to  
 apply this result  to a physical system and show that it leads to certain bounds on the allowed
 values of the physical properties for this system.

\section{Obtaining constraints on the two-particle distribution function } 
Consider the $n$ variable joint probability distribution function $f_n(x_1, x_2, ..., x_n)$ in 
which each variable can take only two values, $-1$ and $1$. The function $f_n(x_1, x_2, ..., x_n)$ is
assumed to be symmetric with respect to the permutation of its variables, 
non-negative everywhere and
normalized. We will use the normalization in which $f_n$ is normalized to one. 
Such  distributions
can describe quite different physical situations. For example, as discussed in more detail below, 
after suitable rescaling $f_n$ can be the joint probability 
distribution of the $z$ spin components
of $n$ particles in a Fermi fluid. It can also describe outcomes of a coin-tossing experiment 
involving $n$ identical coins, or its analogues.
Since we will be dealing with reducing higher order functions to lower order ones it is useful to
introduce parametrizations of such joint probability distributions that will have simple relations 
for the functions of different orders. 
For our purposes a convenient set of parameters is given by the expansion coefficients
of $f_n(x_1, x_2, ..., x_n)$ in terms of products of functions $\phi_0(x_i)$ and $\phi_1(x_i)$, 
that are defined as follows, $\phi_0(x_i)=1$ and $\phi_1(x_i)=x_i$. The expansion of 
a normalized $f_n(x_1, x_2, ..., x_n)$ 
has the following form \cite{Naya}
\begin{widetext}
\be
f_n(x_1, x_2, ..., x_n)=\frac{1}{2^n}\left(1+\sum_{i=1}^n\bra x_i\ket x_i
+\!\sum_{j> i=1}^n\bra x_ix_j\ket x_ix_j 
+\!\!\!\!\sum_{k>j>i=1}^n\bra x_ix_jx_k\ket x_ix_jx_k+ ...+\bra x_1x_2...x_n\ket
 x_1x_2...x_n\right),
\ee
\end{widetext}
where the
expansion coefficients are the moments of $f_n(x_1, x_2, ..., x_n)$ of different orders.
For $f_n(x_1, x_2, ..., x_n)$ with a given $n$, the series terminates at the term involving
the $n^{\rm th}$ order moment. For the functions that are symmetric with respect to 
the permutation of 
any two of their variables, moments of the same order must be the same.
Thus, the normalized and symmetric function $f_n(x_1, x_2, ..., x_n)$ is completely
defined by $n$ parameters, $\bra x_1\ket, \bra x_1x_2\ket,...,\bra x_1x_2...x_n\ket$.
The allowed range of these parameters is obtained from the requirement that
$f_n(x_1, x_2, ..., x_n)\geq 0$ for all coordinate values. Since there are $2^n$ such values
this leads to $2^n$ inequalities.
However, due to the symmetry of $f_n(x_1, x_2, ..., x_n)$ only  $n+1$ of these inequalities 
are independent.  As will be shown 
in the specific examples below, these inequalities specify a closed region of the allowed values 
in the $n$ dimensional  space of
parameters  $\bra x_1\ket, \bra x_1x_2\ket,...,\bra x_1x_2...x_n\ket$.

\begin{figure}
 \includegraphics[width=\columnwidth]{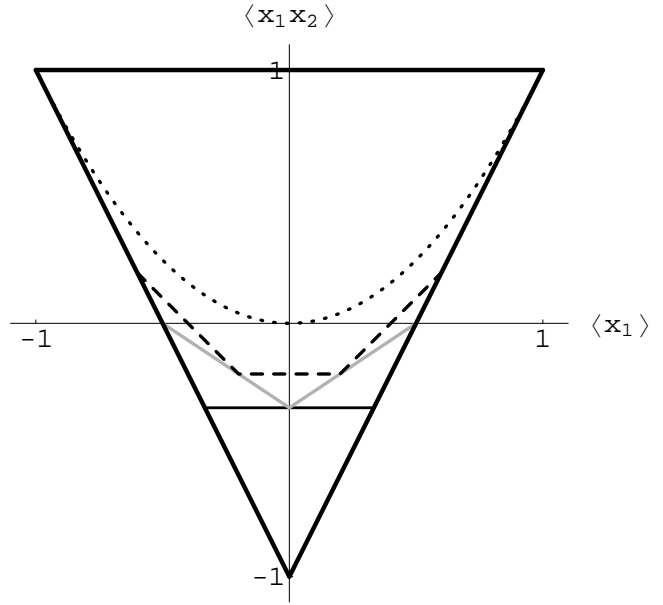}  
 \caption{\label{Graph1}The regions of allowed values for 
 the $f_2$ parameters $\bra x_1\ket$ and $\bra x_1x_2\ket$.
 The region inside the big isosceles triangle is allowed if
 $f_2$ is required to be non-negative, normalized, 
 and permutationally symmetric. 
 The areas of this triangle
 above different lines represent the allowed parameter 
 regions if $f_2$ is obtained from 
 a permutationally symmetric $f_n$. The horizontal solid line 
 corresponds to $n=3$, the gray line to $n=4$, 
 the dashed line to $n=5$,  the dotted parabola to $n\to\infty$.}
 \end{figure}

Let us consider the allowed range of parameters for the function $f_2(x_1, x_2)$ 
provided that it is normalized, symmetric and non-negative. The function 
$f_2(x_1, x_2)$ is completely specified by $\bra x_1\ket$ and $\bra x_1x_2\ket$.
The requirement of the non-negativity leads to the following three inequalities 
\bea
1+2\bra x_{1}\ket+\bra x_{1}x_2\ket&\geq& 0, \non \\
1-2\bra x_{1}\ket+\bra x_{1}x_2\ket&\geq& 0, \non \\
1-\bra x_{1}x_2\ket&\geq& 0.
\eea
If these relations are treated as the equalities, then they define three nonparallel straight lines 
in the $\bra x_1\ket,  \bra x_1x_2\ket$ plane.
The inequalities define a region inside a triangle whose vertices are the intersection points of the 
three straight lines. The $\bra x_1\ket, \bra x_1x_2\ket$ 
coordinates for the vertices of the triangle
are (-1, 1), (0, -1), (1, 1). The region of allowed values is shown on Fig. \ref{Graph1}. 

Now we would like to find out how this region is reduced by the additional
requirement that the function $f_2(x_1, x_2)$ is obtained by summing 
out variable $x_3$ in the distribution 
function $f_3(x_1, x_2, x_3)$ that is  non-negativity, normalized to one, 
and symmetric
with respect to the permutation of any of its three variables.
Such function  $f_3(x_1, x_2, x_3)$ can be completely specified by parameters
$\bra x_1\ket$, $\bra x_1x_2\ket$, and  $\bra x_1x_2x_3\ket$.
The requirement of the non-negativity for $f_3(x_1, x_2, x_3)$ leads to the 
following four
inequalities
\bea
1+3\bra x_1\ket+3\bra x_1x_2\ket+\bra x_1x_2x_3\ket&\geq& 0,\non \\
1-3\bra x_1\ket+3\bra x_1x_2\ket-\bra x_1x_2x_3\ket&\geq& 0,\non \\
1+\bra x_1\ket-\bra x_1x_2\ket-\bra x_1x_2x_3\ket&\geq& 0,\non \\
1-\bra x_1\ket-\bra x_1x_2\ket+\bra x_1x_2x_3\ket&\geq& 0. \label{inequal3}
\eea
These four inequalities define a tetrahedral region  
in the $\bra x_1\ket, \bra x_1x_2\ket, \bra x_1x_2x_3\ket$ space. The vertices of the
tetrahedron have the following $\bra x_1\ket, \bra x_1x_2\ket, \bra x_1x_2x_3\ket$
coordinates, (-1, 1, -1), ( -1/3, -1/3, 1), (1/3, -1/3, -1), (1, 1, 1).

If $f_2(x_1, x_2)$ is obtained from $f_3(x_1, x_2, x_3)$ then the
allowed range of its parameters $\bra x_1\ket$ and $ \bra x_1x_2\ket$ must be
the same that these parameters have in $f_3(x_1, x_2, x_3)$.
This latter range is given by the projection of the tetrahedral  
parameter region for  $f_3(x_1, x_2, x_3)$
onto the $\bra x_1\ket, \bra x_1x_2\ket$ plane.
Since tetrahedron is a convex polytope the projection that we seek 
is either a convex tetragonal or triangular region. If we plot the actual values of the 
$\bra x_1\ket, \bra x_1x_2\ket$ coordinates of the vertices of the tetrahedron 
given above then the tetragonal region is obtained (Fig. \ref{Graph1}). 
Thus, the requirement that $f_2(x_1, x_2)$ is 
obtained from the permutationally symmetric $f_3(x_1, x_2, x_3)$ reduces the range of 
allowed values of  $\bra x_1\ket$ and $ \bra x_1x_2\ket$.

We can apply the same procedure to obtain the restrictions on 
the $f_2(x_1, x_2)$ parameters
if it is obtained from any higher order symmetric $f_n(x_1, x_2,..., x_n)$.
The allowed range of $n$ parameters for $f_n(x_1, x_2,..., x_n)$ is a region
inside a simplex in $n$ dimensions \cite{Munkres}. 
The allowed range of values for $\bra x_1\ket$ and $ \bra x_1x_2\ket$
is determined by projecting this region on the $\bra x_1\ket,  \bra x_1x_2\ket$ plane.
The projection is a convex $(n+1)$-gonal region. The $\bra x_1\ket, \bra x_1x_2\ket$ 
coordinates of the
vertices of the $(n+1)$-gon are obtained from 
the coordinates of the simplex vertices. These latter coordinates are obtained 
from $n+1$ inequalities that specify the simplex region. Inspection of the results 
for several small $n$ values obtained by explicit calculations allows one to come up with 
 the general formulas for the coordinates of the polygon vertices for arbitrary
 $n$,
\be
\bra x_1\ket=\frac{2i-n}{n},\qquad
\bra x_1x_2\ket=\frac{(2 i-n)^2-n}{n(n-1)}, \label{general}
\ee
where $i$ is an integer taking the values from $0$ to $n$.
Examples for $n=4$ and $n=5$ are given on Fig. \ref{Graph1}.

The obtained results allow one to draw certain conclusions about possible
changes of  $f_2(x_1, x_2)$ in an $n$ particle system. Suppose
that the $f_2(x_1, x_2)$ parameters lie in the region that is cut out
by going to an $n+1$ particle system. Then adding an extra particle
to the  $n$ particle system will necessarily change $f_2(x_1, x_2)$. 
If, however,  the $f_2(x_1, x_2)$ parameters lie in the region that is not
affected by going to an $n+1$ particle system then $f_2(x_1, x_2)$
may or may not change depending on the details of the extra particle
addition.

Eqs. (\ref{general}) can be viewed as a parametric form of the function 
$ \bra x_1x_2\ket= \bra x_1x_2\ket(\bra x_1\ket)$.
The explicit form of this function can be obtained by expressing $i$ through $\bra x_1\ket$ from the
first of Eqs. (\ref{general}) and substituting it into the equation for $\bra x_1x_2\ket$. This gives
\be
\bra x_1x_2\ket=\frac{n\bra x_1\ket^2-1}{n-1}. \label{parabola}
\ee
It can be deduced from this  equation that for each $n$ the polygon that 
defines the region of the allowed values  is 
inscribed in a parabola.

For large systems it is of interest to investigate the limit of $n\to\infty$.
It is easy to check using Eqs. (\ref{general}) that both  $\bra x_1\ket$ and
$\bra x_1x_2\ket$ change continuously in the limit $n\to \infty$. 
Taking this limit in Eq. (\ref{parabola})
we obtain 
\be
\bra x_1x_2\ket=\bra x_1\ket^2. \label{limitpar}
\ee
The corresponding region of allowed values is shown on Fig.\ref{Graph1}.
Thus, as $n\to \infty$, the broken  lines formed by $n-2$ sides of each $n$-gone converge
to the parabola given by Eq. (\ref{limitpar}). 

To investigate how fast the areas $A_n$'s of the polygons
converge to their limit at $n\to\infty$ we use the formula for the area of 
a polygon \cite{Beyer} to obtain 
\be
A_n=\frac{4(n+1)}{3n}, \qquad
 \frac{A_n-A_{\infty}}{A_{\infty}}=\frac{1}{n}.
\ee
From the statistical mechanics standpoint this means that if
we want to  
approximate the $n\to\infty$ region by a finite $n$ region 
then we need to consider rather large values of $n$. For example, if
we require a $1\%$ accuracy then $n\geq100$ must be considered.
\section{Effect of constraints on the correlation function}
It is customary in statistical mechanics to
separate the two particle probability distribution into the
uncorrelated and correlated parts as 
\be
f_2(x_1, x_2)=f_1(x_1)f_1(x_2)+g_2(x_1,x_2), \label{cluster}
\ee
where 
\be
f_1(x_1)=\sum_{x_2=\pm1}f_2(x_1, x_2)
\ee
is the one particle distribution function and the correlation function $g_2(x_1,x_2)$ 
is defined by Eq. (\ref{cluster}).
Let us consider how the above constraints on $f_2(x_1,x_2)$ 
affect $g_2(x_1,x_2)$ and $f_1(x_1)$.
The functions $f_1(x_1)$ and $g(x_1,x_2)$ can be written in terms of
parameters $\bra x_1\ket$ and $\bra x_1x_2\ket$ as
\bea
f_1(x_1)&=&\frac{1}{2}\left(1+\bra x_1\ket  x_1\right), \\
 g_2(x_1, x_2)&=&\frac{1}{4}\left(\bra x_1 x_2\ket-\bra x_1\ket^2\right)x_1x_2. \label{correl}
\eea
\begin{figure}
 \includegraphics[width=\columnwidth]{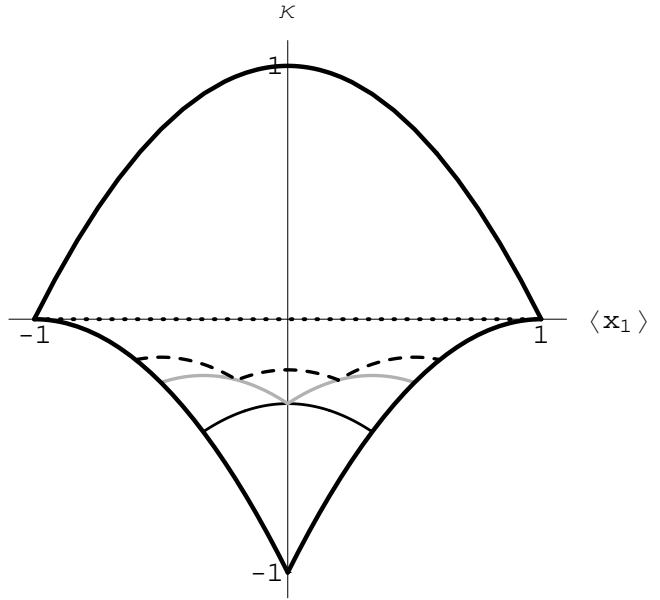}  
 \caption{\label{Graph2}The regions of allowed values for 
  $\bra x_1\ket $ and
 the correlation parameter $\kappa$.
 The region inside the figure formed by the thick solid
  lines is allowed if $f_2$ is required to be non-negative, 
  normalized, and permutationally symmetric. 
 The areas of this figure above different lines represent the 
 allowed parameter regions if $f_2$ is obtained from 
 a permutationally symmetric $f_n$. 
 The thin solid line 
 corresponds to $n=3$, the gray line to $n=4$, 
 the dashed line to $n=5$,  the horizontal  
 dotted line to $n\to\infty$.} 
 \end{figure}
The parameter $\bra x_1\ket $ (that completely specifies $f_1(x_1)$) 
can be viewed as the degree of inhomogeneity 
since $f_1(x_1)$ is constant when $\bra x_1\ket =0$.
The parameter $\kappa=\bra x_1 x_2\ket-\bra x_1\ket^2$ characterizes 
the strength of the correlations.
A surprising result of this analysis is that the line of the zero correlations
on the  $\bra x_1\ket, \bra x_1 x_2\ket$ plane is the same as the limiting parabola
given by Eq. (\ref{limitpar}).
To see more explicitly how the correlation function is affected by the 
constraints, we redraw
 Fig. \ref{Graph1} in the $\bra x_1\ket$,
$\kappa $  space, as shown on Fig. \ref{Graph2}. 

Fig. \ref{Graph2}  shows some interesting relations between 
the correlations and inhomogeneity for finite $n$.
For a fixed $\bra x_1\ket$ the allowed range of $\kappa$ is always a single segment
whereas for a fixed $\kappa$ the allowed range of $\bra x_1\ket$ can be either
a single segment or, for sufficiently negative $\kappa$ a few separate segments.
For this latter case it is impossible to go from one segment to  another without
changing the correlations.
Interestingly, the correlations with the largest negative $\kappa$ require $f_1(x_1)$
to be homogeneous for even $n$ but inhomogeneous for odd $n$.
For $n\to\infty$, only the correlations with $\kappa\geq0$ are allowed.
In this limit, the allowed correlation functions must have $g_2(1,1)\geq0$,
$g_2(-1,-1)\geq0$, and $g_2(1,-1)\leq0$,
$g_2(1,-1)\leq0$

As can be seen on Figs. \ref{Graph1}, \ref{Graph2}, the constraints from
the higher order distribution functions do not affect the maximum possible range of 
$\bra x_1\ket$. This is to be expected since $\bra x_1\ket$ is
the only parameter characterizing $f_1$ and for every physical $f_1$
we can construct a permutationally symmetric $f_n$ as a product
of $n$ $f_1$'s.

\section{Bounds on the magnetization and magnetic susceptibility
in spin $1/2$ quantum fluids}
As an example of application of our results to a  physical system
we will investigate allowed ranges of average $z$ component of the total
spin $\bra S^z\ket$ and its variance $\bra( S^z)^2\ket-\bra S^z\ket^2$
in a Fermi fluid \cite{Huang} or Bose fluid with (iso)spin $1/2$ \cite{Nikuni}.
Here
\be
S^z=\sum_i s^z_i =\frac{\hbar}{2}\sum_i\sigma^z_i \label{spin}
\ee
 and $s^z_i$ is the $z$ component of spin for particle $i$ and 
 $\sigma^z_i$ is the corresponding Pauli matrix.
The quantities $\bra S^z\ket$ and  $\bra( S^z)^2\ket-\bra S^z\ket^2$ are of
interest because for a wide range of   systems in equilibrium
they are proportional to the magnetization and magnetic susceptibility, respectively.    
Note, however, that our results for $\bra S^z\ket$ and $\bra( S^z)^2\ket-\bra S^z\ket^2$
apply to arbitrary states, not necessarily equilibrium ones. Since $\sigma^z_i$
has the eigenvalues of $\pm 1$ we can see that function $f_n$ discussed above
can be viewed as  the probability distribution of the $\sigma^z_i$ eigenvalues for $n$
particles with variable  $x_i$ denoting eigenvalues of  $\sigma^z_i$.

Before proceeding with our analysis we need to be sure that the quantum nature
of the Fermi or Bose fluid does not impose additional restrictions on $f_n$. This is 
indeed the case. We will not go into details of the proof here.
The outline of the proof
is as follows. 
It  can be shown that the vertices of $n$ dimensional
simplex domains for the allowed $f_n$ parameters correspond to
the total density matrices composed of the wave functions that are eigenstates
of $S^z$ with a given eigenvalue. (For an $n$ particle  system there are 
$n+1$ different $S^z$ eigenvalues, as many as there are vertices.) 
Since the density matrices for the 
vertices exist and since any simplex domain is convex, the density matrices for
the points inside the simplex also exist. Thus, $\sigma^z_i$'s can be rigorously
treated as classical variables as far as the joint distributions of their 
eigenvalues are concerned.

\begin{figure}
 \includegraphics[width=\columnwidth]{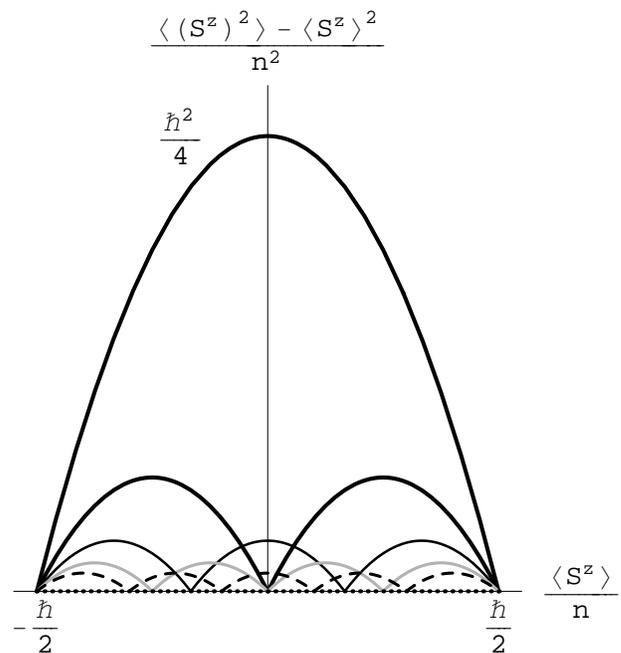}  
 \caption{\label{Graph3}The regions of allowed values for 
 $\bra S^z\ket/n$ and $\left(\bra( S^z)^2\ket-\bra S^z\ket^2\right)/n^2$.
 The regions bounded from above by the big parabolic arc 
 and from below by series of  $n$ smaller arcs drawn in different styles
 are allowed for the $n$ particle Fermi fluid.
 The thick solid line 
 corresponds to $n=2$, the thin solid line to $n=3$, 
 the gray line to $n=4$,  the dashed line to $n=5$, the dotted straight line to $n\to\infty$.}
 \end{figure}

Using Eq. (\ref{spin}) and replacing $\sigma^z_i$ with $x_i$ we obtain
for the average $S^z$  and its variance.
\bea
\bra S^z\ket&=&\frac{\hbar}{2}n\bra x_1\ket \non \\
& &\non \\
\bra (S^z)^2\ket-\bra S^z\ket^2
&=&\frac{\hbar^2}{4}\left(n(n-1)\bra x_1x_2\ket -n^2\bra x_1\ket^2+n\right) \non \\
\eea
Using the allowed range of parameters for $\bra x_1\ket$ and $\bra x_1x_2\ket$
we can obtain the allowed domains for $\bra S^z\ket$ and  $\bra( S^z)^2\ket-\bra S^z\ket^2$.
The results are shown on Fig. \ref{Graph3}.  
 Qualitative physical interpretation of these results  is straightforward. 
 Each point of the allowed domain with $\bra( S^z)^2\ket-\bra S^z\ket^2=0$ correspond to 
 the total density matrices composed of eigenstates of $S^z$ with the same eigenvalue, i. e.
  $\bra S^z\ket$ is a multiple of $\hbar/2$. For $\bra S^z\ket$ that is not a multiple of $\hbar/2$
  $\bra( S^z)^2\ket-\bra S^z\ket^2$ is necessarily greater than zero (because not all spins are
  either up or down). Generally, for smaller $\bra S^z\ket$ values the allowed range of 
  $\bra( S^z)^2\ket-\bra S^z\ket^2$ is larger.

Consider a Fermi fluid whose Hamiltonian involves the coupling to the magnetic field ${\bf { \cal H}}$
of the form $H_m=-\frac{e}{mc}S^z{\cal H}$. For the system in the thermal equilibrium 
the magnetization per unit volume, $M$
is given by
\be
M=\frac{e}{mcV}\bra S^z\ket
\ee
If $H_m$ commutes with the total Hamiltonian then the magnetic susceptibility, $\chi$ is given by 
\be
\chi=\frac{\beta}{V}\left(\frac{e}{mc}\right)^2\left(\bra (S^z)^2\ket-\bra S^z\ket^2\right)
\ee
Thus, with suitable rescaling, Fig. \ref{Graph3} describes the allowed domains of $M$
and $\chi$ for any model satisfying the above requirements. As can be expected, no such model
can be diamagnetic ($\chi$ is always $\geq 0$). 
\section{Concluding remarks}
The generalization of our analysis to the functions whose variable 
can take more than
two values is straightforward.
If each variable of $f_2(x_1,x_2)$ is allowed to take $l$ values,
then $f_2(x_1,x_2)$ can be  completely specified by $(l(l+1)-2)/2$ parameters.
Parameters of the higher order functions can always be chosen in such a way
that they include $(l(l+1)-2)/2$ parameters of $f_2(x_1,x_2)$. Constraints on the
latter parameters are obtained by projecting the allowed parameter range of the
higher order function on the $f_2(x_1,x_2)$ parameter subspace. 

The case of continuous coordinates is more complicated since infinitely
many parameters are needed to specify the distribution functions. In this case,
our approach can be used to investigate the finite dimensional subspaces of 
the $f_2(x_1,x_2)$ parameter space in some regions of interest, e. g., in the
vicinity of functions with a certain characteristic wavelength.

\begin{acknowledgments}
The author would like to thank Eric. R. Bittner
for useful discussions.
This work was funded in part through grants from the National Science
Foundation and the Robert A. Welch foundation.
\end{acknowledgments}

\end{document}